\documentclass[prd,twocolumn,showpacs,showkeys,preprintnumbers,floatfix,
nofootinbib,superscriptaddress]{revtex4}
\usepackage{amsfonts} 
\usepackage{amssymb} 
\usepackage{amsmath} 
\usepackage{graphicx} 
\usepackage{subfigure} 
\usepackage{array} 
\usepackage{dcolumn} 
\usepackage{bm} 
\usepackage{latexsym} 
\usepackage{longtable} 
\usepackage[hypertexnames=false]{hyperref} 
\graphicspath{{./figs/}}
\renewcommand{\tilde}{\widetilde}
\begin{document}


\title{Energy Loss Calculations of Moving Defects for General Holographic Metrics}
\author{John F. Fuini III}
\email[Email: ]{fuini@u.washington.edu}
\author{Andreas Karch}
\email[Email: ]{akarch@uw.edu}

\affiliation{Physics Department, University of Washington,
 Seattle, WA 98195-1560, USA \\
}

\date{\today}
\begin{abstract}
We extend the ideas of using AdS/CFT to calculate energy loss of extended defects in strongly coupled systems to general holographic metrics.  We find the equations
of motion governing uniformly moving defects of various dimension and determine the corresponding energy loss rates in terms of the metric coefficients.
We apply our formulae to the specific examples of both bulk geometries
created by general D$p$-branes, as well as to holographic superfluids. For the D$p$-branes, we find that the energy loss of our
defect, in addition to the expected quadratic dependence on velocity, depends on velocity only via an effective blueshifted temperature - despite the existence of a microscopic length scale in the theory.
We also find, for a certain value of $p$ and dimension of the defect, that the energy loss has no dependence on temperature or velocity other than
the aforementioned quadratic dependence on velocity.  For the superfluid example, we find agreement with previous results on the existence of a cutoff velocity, below which the probe
experiences no drag force.  For both examples we can easily extend the equations of motion and energy loss to
defects of larger dimension.
\end{abstract}
\keywords{AdS/CFT}
\maketitle

\section{Introduction \label{sec:intr}}

The duality between Type IIB string theory on $AdS_5$ $\times$ $S^5$ and ${\cal N} = 4$ SU($N$) supersymmetric Yang-Mills (SYM) \cite{Maldacena:1997re,Gubser:1998bc,Witten:1998qj} has been
studied extensively as a means to provide insight to the inner workings of strongly coupled
systems where perturbation theory is not valid, and where lattice gauge theory and Monte Carlo techniques are available but
struggle with real time physics.  Motivated by the QGP created at RHIC,
holography has helped guide our understanding of shear viscosity, drag, screening
length and jet quenching to name a few; see \cite{Karch:2011ds} for a recent perspective on these developments by one of us.  The dual description of a classical string
ending on a probe brane has given information on the characteristics of a point particle,
possibly a quark, traveling through the Yang-Mills plasma, which is thought to be
a good approximation for the strongly coupled QCD.  Recently it has been seen that one
can also study extended defects with dimensions larger than a point, living in $AdS_d$
spaces, the idea being that one can shed new light on the dynamics of energy loss at strong coupling~\cite{stefan}.
We extend the ideas of~\cite{karchquarkdrag, herzogquarkdrag, stefan} to a general metric with a compact internal space.
 We produce general solutions for the equation of motion and the energy loss of an extended defect moving
uniformly through the bulk whose geometry is described by a generic brane-like metric.
We study two examples in detail, general D$p$-brane metrics and holographic superfluids.  After working out the
energy loss for D$p$-branes, we find that the energy loss of these extended probes is given by simple power laws in
velocity and temperature, revealing that the energy loss depends only on an effective temperature multiplied by the velocity squared
- the moving probe being affected only by a blue-shifted
energy density - despite the existence of an intrinsic scale in the underlying theory.
We then apply our results to dragging string- and sheet-like defects through holographic superfluids.

We organize these ideas as follows: In section II, we will calculate a general equation of motion and solution to a uniformly
moving defect in a general bulk theory.  We then find a general energy loss formula for said defect.  In section III,
we apply these results to metrics created by general D$p$-branes.  In section IV, we apply our results
to a holographic superfluid.

\section{Calculations \label{calculations}}

We want to study a general holographic brane-like metric.  This metric will preserve the symmetries
of the dual field theory. Therefore, it should be rotationally
invariant and translationally invariant in both the spatial and time dimensions.  In addition it should preserve the
isometries of the internal space.  To these ends, we introduce the following diagonal metric
\begin{eqnarray}
ds^2=G_{\mu\nu}dx^{\mu}dx^{\nu}
\nonumber\\
=G_{tt}dt^2+G_{xx}\sum_{i=1}dx_i^2+G_{uu}du^2+G_{\theta\theta}d\Omega^2,
\label{generalform}
\end{eqnarray}
with $N+2$ total space-time dimensions.
At this point, the $N+1$ spatial coordinates are arbitrarily
separated into $M$ Cartesian coordinates and $N-M$ spherical angles of the internal space. The additional radial coordinate is
denoted by $u$. It is
understood that $G_{\theta\theta}$ is only a function of $u$ and since none of our results depend on the details of the
internal space we will take $d\Omega^2$ be a unit sphere - i.e. $d\Omega^2$ contains the appropriate terms for the
angular variables of a unit sphere of arbitrary dimension. We are only considering metrics whose $G_{ii}$ depend
only on $u$, and whose $G_{tt}$ and $G_{xx}$ grow with $u$ at the same rate, both of which grow faster with $u$ than
$G_{\theta\theta}$.  The rationale behind this is that the gravity theory in the bulk has a dual interpretation in terms
of a field theory at the $u = \infty$ boundary which lives in $M+1$ space-time dimensions.

In a manner similar to~\cite{stefan}, we introduce a defect of spatial dimension $n+1$ in the bulk. 
The defect has $n$ spatial dimensions orthogonal to $u$ which 
are divided into $m$ infinitely extended Cartesian
directions on the boundary, denoted $\vec{y}$, and $n-m$ angular directions, denoted $\vec{\theta}$.  
It will then move in an additional transverse direction, $x$.
Since the dimensionality of the bulk puts constraints on the size of the defect we find $m+1\leq M$ and $n+1\leq N$.

In the static gauge, the world-volume map is of the form $X =(t,u,\vec{y}, \vec{\theta},x(t,u,\vec{y},\vec{\theta}),
\vec{z}=const)$ where $\vec{z}$ are any additional unused orthogonal coordinates in the bulk.   The induced metric on the world volume is defined
as $g_{\alpha\beta} = G_{\mu\nu}\frac{\partial X^{\mu}}{\partial\sigma^{\alpha}}\frac{\partial
X^{\nu}}{\partial\sigma^{\beta}}$. We find $g = \det (g_{\alpha\beta})$ to be
\begin{eqnarray}
\label{inducedg}
g = G_{xx}^m G_{\theta\theta}^{n-m} [G_{tt}G_{uu}+G_{tt}G_{xx} (x_{,u})^2
\nonumber\\
+ G_{uu}G_{xx} (x_{,t})^2 + G_{tt}G_{uu}\sum_{i}^m (x_{,y_i})^2
\nonumber\\
+\frac{G_{tt}G_{uu}G_{xx}}{G_{\theta\theta}}\sum_{j}^{n-m} (x_{,\theta_j})^2]
\end{eqnarray}
where $(x_{,\sigma})$ denotes the partial derivative of $x$ with respect to the coordinate $\sigma$.

Our general metric will have some intrinsic scale governed by its radii of curvature, all of which we take parametrically to be of the same
order $R$. The bulk theory will be governed
by classical gravity if $R$ abides by $M_{pl}R\gg1$, where $M_{pl}^{N} = 1/16\pi G$,
  (ensuring our curvature is not too large in Planck units). We want to describe the defect by the following Nambu-Goto-like
  action
\begin{equation}
\label{nglike}
S = -T_0 \int e^{-\Phi}\sqrt{-g} \prod_i d\sigma_i
\end{equation}
where $g$ is given by (\ref{inducedg}), $T_0$ is the
tension and the integral is over all world-volume coordinates. $\Phi$ is a function of the background scalar fields, the dilaton field,
$\phi$, for example. Unlike the coupling to the background metric, the functional form of the coupling to the background scalars is not fixed by diffeomorphism invariance and in principle $e^{-\Phi}$ can be a complicated function of the background scalars. As long as the background scalars respect the symmetries of the metric, they can only depend on the radial coordinate $u$ and we can treat $\Phi$ as a function of $u$. Two important probes we are going to consider in examples for $\Phi(u)$ are a fundamental string, whose action has the form (\ref{nglike}) with $\Phi=0$, and probe D-branes, whose action is of the form (\ref{nglike}) with $\Phi=\phi$.
We can trust the classical treatment of this action
so long as $T_0 R^{n+2}\gg 1$. Additionally, demanding that 
$(M_{pl}R)^N\gg T_0R^{n+2}$ will render gravitational back-reaction negligible.

With $\mathcal{L}=-T_0 e^{-\Phi}\sqrt{-g}$ we find the following canonical momentum densities:
\begin{eqnarray}
\label{momenta}
\Pi_x^t=\frac{\partial \mathcal{L}}{\partial (x_{,t})}=-T_0
\frac{G_{uu}G_{xx}^{m+1}G_{\theta\theta}^{n-m}}{\sqrt{-g}}(x_{,t})e^{-\Phi}
\nonumber\\
\Pi_x^u=\frac{\partial \mathcal{L}}{\partial (x_{,u})}=-T_0
\frac{G_{tt}G_{xx}^{m+1}G_{\theta\theta}^{n-m}}{\sqrt{-g}}(x_{,u})e^{-\Phi}
\nonumber\\
\Pi_x^{y_i}=\frac{\partial \mathcal{L}}{\partial (x_{,y_i})}=-T_0
\frac{G_{tt}G_{uu}G_{xx}^{m}G_{\theta\theta}^{n-m}}{\sqrt{-g}}(x_{,y_i})e^{-\Phi}
\nonumber\\
\Pi_x^{\theta_i}=\frac{\partial \mathcal{L}}{\partial (x_{,\theta_i})}=-T_0
\frac{G_{tt}G_{uu}G_{xx}^{m+1}G_{\theta\theta}^{n-m-1}}{\sqrt{-g}}(x_{,\theta_i})e^{-\Phi}
\end{eqnarray}

Requiring a vanishing variation in our action yields
\begin{equation}
\sum_{i}^n \partial_{\sigma_i} (\Pi_{\mu}^{\sigma_i}) = 0
\end{equation}

Setting $\mu$ to x gives us our equation of motion.

We are interested in a solution of a uniformly moving object.  The defect should move in a direction transverse to
its spatial extent and travel with a constant velocity in the x direction.  Thus, our ansatz is
\begin{equation}
\nonumber x(t,u,\vec{y})= vt + x(u).
\end{equation}

With this form, $g$ becomes independent of time and our equation of motion, $\sum_{i}^n \partial_{\sigma_i} \Pi_{mu}^{\sigma_i} = 0$, will then reduce to
\begin{equation}
\nonumber\partial_u \left( \frac{G_{tt}G_{xx}^{m+1}G_{\theta\theta}^{n-m}}{\sqrt{-g}}(x_{,u})e^{-\Phi} \right)=0,
\end{equation}
which gives us
\begin{equation}
\label{xp}
(x_{,u})^2 = \frac{C^2e^{2\Phi}(-g)}{G_{tt}^2G_{xx}^{2(m+1)}G_{\theta\theta}^{2(n-m)}},
\end{equation}
and plugging in for $g$
\begin{equation}
\label{xonsol}
\textstyle(x_{,u})^2 = -(Ce^{\Phi})^2\frac{G_{uu}}{G_{tt}G_{xx}e^{2\Phi}}\{\frac{G_{tt}+G_{xx}v^2}{G_{tt}G_{xx}^{m+1}G_{\theta\theta}^{n-m}e^{-2\Phi}+C^2}\}
\end{equation}

We expect on physical grounds, that $(x_{,u})$ should be real, and thus $(x_{,u})^2$ and $(-g)$ should be positive.  From
(\ref{xonsol}), we see that this will be true if $G_{tt}+G_{xx}v^2$
and $G_{tt}G_{xx}^{m+1}G_{\theta\theta}^{n-m}e^{-2\Phi}+C^2$ both switch sign at the same value of $u$.  Assuming that
there is at most one root, call it $u_c$, we can solve for it from
\begin{equation}
\label{uccond}
 G_{tt}(u_c)+G_{xx}(u_c)v^2=0
\end{equation}
and if this root exists, we have
\begin{equation}
\label{c}
\nonumber\\C = \pm e^{-\Phi}\sqrt{-G_{tt}G_{xx}^{m+1}G_{\theta\theta}^{n-m}}|_{u=u_c}.
\end{equation}
The case where $u_c$ does not exist occurs in our example of a holographic superfluid and will be discussed in section
IV. The induced metric can be diagonalized and this diagonal form has a time component which vanishes when $G_{tt}(u_c)+G_{xx}(u_c)v^2=0$.  
This tells us that $u_c$ denotes the worldvolume horizon.

Using (\ref{uccond}) to plug in for $G_{tt}$ and defining $\tilde{C} = C/v$ we have
\begin{equation}
\label{Ctilde}
\tilde{C} = \pm e^{-\Phi}\sqrt{G_{xx}^{m+2}G_{\theta \theta}}|_{u=u_c}
\end{equation}

The momentum loss rate, due to momentum flowing along the defect and towards the horizon, is given by -$\Pi_x^u$
which is seen from (\ref{momenta}) and (\ref{xp}) to be
\begin{equation}
\label{pitu}
-\Pi_x^u = T_0C.
\end{equation}
The momentum loss rate directly gives us the drag force density. The energy loss rate, $-\Pi_t^u$, is simply $v$ times the momentum loss rate.
Physically, we expect that we have energy flowing towards
the horizon of the black brane, which requires our loss rate to be positive. We thus pick the positive sign for $\tilde{C}$.
\begin{equation}
\label{cagain}
\tilde{C} = e^{-\Phi}\sqrt{G_{xx}^{m+2}G_{\theta\theta}^{n-m}}|_{u=u_c}.
\end{equation}
While we have established that this stationary solution is a consistent solution
to the equations of motion, what is less obvious is that it is stable. Small
fluctuations around dragging sheets in AdS$_7$ have recently been studied
in \cite{Janiszewski:2011sx}
and it has been found that these fluctuations do not exhibit any instabilities,
that is any modes that grow exponentially in time.
A new potential instability in our case is the slipping mode on the internal space,
that is fluctuations in the $\theta$ directions that take our defect off the equator of the internal
sphere. As such fluctuations reduce the volume of the defect (and hence its
potential energy), they clearly correspond to negative mass squared modes and are hence potentially
problematic. It is well known from the case of static defects, starting with the work
on flavor probe branes \cite{Karch:2002sh}, that these negative mass squared slipping modes
often are actually stable. For a background AdS space the basic physics behind this is 
the BF bound \cite{Breitenlohner:1982bm}. The potential energy gain of the fluctuation is
offset by the kinetic energy cost of any fluctuation in a spacetime geometry that 
effectively corresponds to a finite size box. A similar effect also occurs in the more general
holographic metrics. In particular, it has been shown in \cite{Camino:2001at}
that supersymmetric D$q$-brane
defects in black D$p$-brane backgrounds (which will be the first example we apply our results to)
have stable slipping modes. For non-supersymmetric defects stability of the slipping mode will
have to be checked on a case by case basis.

\section{p-branes \label{p-branes}}

We now turn to the example of general D$p$-branes which create geometries dual to SYM in $p+1$ dimensions
on the boundary, at finite temperature. From~\cite{pbranes} we see that in the limit that
\begin{equation}
g^{2}_{YM}=(2\pi)^{p-2}g_s\alpha'^{(p-3)/2}=\mathsf{fixed}, \nonumber
\end{equation}
as
\begin{equation}
\alpha'\rightarrow 0 \nonumber
\end{equation}
where $g_s = e^{\phi_{\infty}}$, and $g_{YM}$ is the Yang-Mills coupling constant,
the D$p$ brane metric becomes,
\begin{eqnarray}
ds^2=\alpha'\{\frac{u^{(7-p)/2}}{g_{YM}\sqrt{d_pN'}}\left[-(1-\frac{u_0^{7-p}}{u^{7-p}})dt^2+dy_{\parallel}^2\right]
\nonumber\\
+\frac{g_{YM} \sqrt{d_pN'}}{u^{(7-p)/2}(1-\frac{u_0^{7-p}}{u^{7-p}})}du^2
\nonumber\\
+g_{YM}\sqrt{d_pN'}u^{(p-3)/2}d\Omega_{8-p}^2\}
\label{dpmetric}
\end{eqnarray}
which indeed is of the general form (\ref{generalform}).  Here, $N'$ is the number of branes and 
 $d_p = 2^{7-2p} \pi^{\frac{9-3p}{2}}\Gamma(\frac{7-p}{2})$. 

The dilaton is
\begin{equation}
e^{\phi}=(2\pi)^{2-p}g^2_{YM}(\frac{g^2_{YM}d_pN'}{u^{7-p}})^{\frac{3-p}{4}},
\end{equation}
and
\begin{equation}
u_0^{7-p}= \frac{\Gamma(\frac{9-p}{2})2^{11-2p}\pi^{\frac{13-3p}{2}}}{(9-p)}g^4_{YM}\epsilon,
\end{equation}
where $\epsilon$ corresponds to the energy density of the Yang-Mills theory.

Following the outline laid out in the previous section, we first need to find $u_c$ from (\ref{uccond}).  Extracting $G_{tt}$
and $G_{xx}$ from (\ref{dpmetric}), and plugging into (\ref{uccond}) we find
\begin{equation}
u_c = \frac{u_0}{(1-v^2)^{1/(7-p)}}
\end{equation}

At this stage we need to commit to the nature of the probe, that is we need to chose a particular function $\Phi$.
Let us first focus on the case where the dragging object is a D$(n+1)$-brane itself,
 in which case it couples to the string theory dilaton $\phi$ with an overall prefactor of $e^{-\phi}$ in the action, that is $\Phi=\phi$.
From (\ref{pitu}) we find we can isolate the dependence of $\tilde{C}$ on $v$ and $T$
\begin{equation}
\label{defineab}
\tilde{C} = P(1-v^2)^A T^B
\end{equation}
Where the prefactor is,
\begin{equation}
P=\left(\frac{(\alpha')^{n+2}(2\pi)^{2(p-2)}}{g_{YM}^{4}(g_{YM}\sqrt{d_pN})^{2m-n-p+5}}\right)^{\frac{1}{2}}\left(\frac{4\pi}{7-p}\right)^{B},
\end{equation}
\begin{equation}
A = -\frac{1}{4}\left(5+m-p+\frac{(p-3)(n-m)}{(7-p)}\right)
\end{equation}
and
\begin{equation}
B= - 2\frac{(7-p)}{(5-p)}A.
\end{equation}
Here we have made use of the relation
\begin{equation}
\label{u0}
u_0 = \left(\frac{4\pi T}{7-p}\right)^{\frac{2}{5-p}}.
\end{equation}

Various dependencies of $\tilde{C}$ on velocity and temperature are shown in the following tables for different values of $p$ (rows) and
$n$ (columns). Each table has a specific value of $m$, and the entries in the table are of the form \{$A$,$B$\}, where $A$
and $B$ are defined as in (\ref{defineab}).

Table I shows the dependencies for allowing our defect to have a point on the internal sphere, $n=m$.
        \renewcommand{\arraystretch}{1.3}
	\begin{table}
	\begin{tabular}{ | l | p{1.4cm} | p{1.4cm} | p{1.4cm} | p{1.4cm} | p{1.4cm} |}
	\hline
	$p \backslash n$   & 0 & 1 & 2 & 3 & 4 \\ \hline
	1 & $\{-1,3\}$  & $\{-\frac{5}{4},\frac{15}{4}\}$ & N/A & N/A & N/A\\ \hline
	2 & $ \{-\frac{3}{4},\frac{5}{2}\}$ & $\{-1,\frac{10}{3}\}$ & $\{-\frac{5}{4},\frac{25}{6}\}$ & N/A & N/A\\ \hline
	3 & $ \{-\frac{1}{2},2\}$ & $\{-\frac{3}{4},3\}$ & $\{-1,4\}$ & $\{-\frac{5}{4},5\}$ & N/A\\ \hline
	4 & $ \{-\frac{1}{4},\frac{3}{2}\}$ & $\{-\frac{1}{2},3\}$ & $\{-\frac{3}{4},\frac{9}{2}\}$ & $\{-1,6\}$ & $\{-\frac{5}{4},\frac{15}{2}\}$\\ \hline
	\end{tabular}
      \caption{\{$A$,$B$\} displayed for the case where the defect does not extend into the internal space, $m=n$.}
	\end{table}
For table II we allow one spatial dimension of the defect to go to the internal space (our defect must have a minimum dimension of $n=1$, as we are insisting
that one of our spatial dimensions lives in the internal sphere). For table III we allow two spatial dimensions of our defect to go to the internal space (now our defect must have at
least 2 spatial dimensions, $n=2$, and $p$ must contain $n$, meaning $p \geq 2 $).
\renewcommand{\arraystretch}{1.3}
	\begin{table}
	\begin{tabular}{ | l | p{1.4cm} | p{1.4cm} | p{1.4cm} | p{1.4cm} |}
	\hline
	$p \backslash n$    & 1 & 2 & 3 & 4 \\ \hline
	1 & $ \{-\frac{11}{12},\frac{11}{4}\}$ & N/A & N/A & N/A \\ \hline
	2 & $ \{-\frac{7}{10},\frac{7}{3}\}$ & $\{-\frac{19}{20},\frac{19}{6}\}$ & N/A & N/A \\ \hline
	3 & $ \{-\frac{1}{2},2\}$ & $\{-\frac{3}{4},3\}$ & $\{-1,4\}$ & N/A \\ \hline
	4 & $ \{-\frac{1}{3},2\}$ & $\{-\frac{7}{12},\frac{7}{2}\}$ & $\{-\frac{5}{6},5\}$ & $\{-\frac{13}{12},\frac{13}{2}\}$ \\ \hline
	\end{tabular}
	\caption{\{$A$,$B$\} displayed for the case where the defect extends into one dimension in the internal space, $m=n-1$.}
	\end{table}

\renewcommand{\arraystretch}{1.3}
	\begin{table}
	\begin{tabular}{ | l | p{1.4cm} | p{1.4cm} | p{1.4cm} | }
	\hline
	$p \backslash n$   & 2 & 3 & 4 \\ \hline
	2 & $ \{-\frac{13}{20},\frac{13}{6}\}$ & N/A & N/A  \\ \hline
	3 & $ \{-\frac{1}{2},2\}$ & $\{-\frac{3}{4},3\}$ & N/A  \\ \hline
	4 & $ \{-\frac{5}{12},\frac{5}{2}\}$ & $\{-\frac{2}{3},4\}$ & $\{-\frac{11}{12},\frac{11}{2}\}$  \\ \hline
	\end{tabular}
	\caption{\{$A$,$B$\} displayed for the case where the defect extends into two dimensions of the internal space, $m=n-2$.}
	\end{table}

The D$p$-brane energy density obeys the following relation, $\epsilon \sim T^{2\frac{(7-p)}{5-p}}$. Under
a boost, we expect $\epsilon \rightarrow \gamma^2\epsilon$ and thus
 $T^{2\frac{(7-p)}{(5-p)}} \rightarrow \gamma^2 T^{2\frac{(7-p)}{(5-p)}}$.  This motivates the definition of
\begin{equation}
  T_{\mathsf{eff}} = \gamma^\frac{(5-p)}{(7-p)} T = \gamma^{-2A/B} T.
\end{equation}
With this definition of the effective temperature we see that, analogous to relation (\ref{u0}),
we have an identical relation between the worldvolume horizon and the effective temperature,
\begin{equation}
\label{uc}
u_c = \left(\frac{4\pi T_{\mathsf{eff}}}{7-p}\right)^{\frac{2}{5-p}}.
\end{equation}
Consequently we see that $T^B =
 (1-v^2)^{-A} T_{\mathsf{eff}}^B$ and so we can rewrite $\tilde{C}$ as
\begin{equation}
 \tilde{C} = P T_{\mathsf{eff}}^B.
\end{equation}
This tells us that the loss rate of the moving defect is only dependent on velocity in a trivial way -  the defect only sees a blueshifted
energy density - and that there are no sensitivities to the microscopic details of the plasma despite the fact that $g_{YM}$
is a dimensionfull quantity and hence defines a microscopic scale in the system. Presumably this is a consequence of the hidden conformal invariance that is present in the D$p$ brane systems as first exhibited in 
\cite{Kanitscheider:2009as}.

There is an area of overlap between this work and~\cite{stefan}, where in the latter, various dimensional defects are
studied is AdS spaces of variable dimension.  Our results for a D$p$-brane with $p=3$ reproduce the equation of motions
and loss rates found in~\cite{stefan} for the case $AdS_5$, as it should. For the case of a pointlike
defect, $m=n=0$, our results can be compared to the formulas quoted for the dragging string in \cite{Karch:2009eb} following the analysis of dragging strings in general holographic metrics performed in \cite{herzogquarkdrag}. Our $m=n=0$ results are for a dragging D-string, as we included an overall $e^{-\phi}$ coupling in the action.
To compare with the results for the dragging fundamental string we have to set $\Phi=0$ in our analysis (that is, the worldvolume action is independent of the dilaton). It is easy to see that in this case our general expression (\ref{pitu}) indeed nicely reduces to the result of \cite{Karch:2009eb}.
Other than the trivial velocity dependence, the energy loss rate still only depends on velocity and temperature via a power of $u_c$,
and hence, due to (\ref{uc}), via the effective temperature.

Last but not least, it is interesting
to note that for the case $m=n=1$ and $p=6$, our defect's energy loss rate is independent of both velocity and temperature, other than the
expected velocity squared dependence. In $p=6$ case, there is no good decoupling limit~\cite{pbranes} and it is not clear what significance should be attached to this
result.

\section{Superfluidity}
Pointlike probes have been used to study superfluids that have a gravity dual \cite{Gubser:2008px,Hartnoll:2008vx}. This is an area in
which strongly interacting extended defects exist in nature (vortices in Liquid Helium) and thus suggests a
possible analysis using a gauge-string
duality. Following the layout of~\cite{superfluid}, we are interested in using a superconducting black hole in $AdS_5$.

The bulk theory has metric, gauge field, and a complex scalar field (magnitude $\eta$ and phase $\theta$) degrees of
freedom, and is governed by
\begin{equation}
\mathcal{L}_{bulk}=R-\frac{1}{4}F_{\mu\nu}^2 -
\nonumber\\\frac{1}{2}\left[(\partial_{\mu}\eta)^2+\Sigma(\eta)(\partial_{\mu}\theta-q A_{\mu})^2\right]-V(\eta).
\end{equation}
This Lagrangian density allows a charged black brane solution to the metric of the form
\begin{equation}
\label{blackbrane}
ds^2 = e^{2A(u)}(-h(u)dt^2+d\vec{x}^2) + \frac{du^2}{h(u)},
\end{equation}
where $A_{\mu}dx^{\mu} = A_0(u)dt$, $\eta = \eta(u)$, $\theta = 0$, $A(u)$ is the warp factor, $h(u)$ is the blackening function,
and u is the ``radial coordinate" that is defined between $-\infty$ and $\infty$. $V$ and $\Sigma$ are in principle
free functions of $\eta$ which in reference \cite{superfluid} are taken to be $V(\eta)= -\frac{3}{L^2}\cosh^2(\frac{\eta}{2})(5-\cosh(\eta))$ and
$\Sigma(\eta)=\sinh^2(\eta)$. These particular forms are
required for a consistent truncation of Type II B supergravity on a Sasaki-Einstein 
manifold \cite{Gubser:2009qm}. The blackening
function smoothly interpolates from 1 at large $u$, to its asymptotically value $v_{IR}^2$ as $u \rightarrow -\infty$.

We can now apply our general formulae to the gravity dual metric for this holographic superfluid.
We first reproduce some results of \cite{superfluid}. A string in the bulk has the following action,
\begin{equation}
\label{defectaction}
S = -\int d\sigma d\tau \frac{1}{2\pi\alpha'}Q(\eta)\sqrt{-g},
\end{equation}
where $Q = \cosh(\frac{\eta}{2})$, and $\alpha'$ is the square of the string length
scale and goes to zero in the limit of infinite string tension.
We compare their action for the string (20) to our general formula (\ref{nglike}) setting $m=n$ and
$n=0$ so that we are
discussing the same defect.  We find that we should make the associations $T_0 \rightarrow \frac{1}{2\pi\alpha'}$,
and $e^{-\Phi(u)}\rightarrow Q(\eta(u))$.

Following our prescription for finding the solution to a uniformly moving defect,
we first find the root of equation (\ref{uccond})
where we are now using the metric appropriate for our bulk theory in $AdS_5$ (\ref{blackbrane}).  We see that
$G_{tt}=-e^{2A}h$, $G_{xx} = e^{2A}$ and $G_{uu}=h^{-1}$.

From (\ref{uccond}), $u_c$ should be given by $h(u_c) = v^2$. It is clear from the form of $h(u)$ that if $v^2<v_{IR}^2$
there is no solution, and thus the value $h(u)$ approaches as $u\rightarrow -\infty$ defines a cutoff velocity,
 $v_{IR}$~\cite{superfluid}.  Defects whose velocities are below this cutoff experience zero drag force.

 For velocities above the cutoff, we find the non-zero drag force density from the momentum
 density of our uniformly moving defect, $\Pi_x^u = -T_0 \tilde{C} v$, which comes from (\ref{momenta}), with $(x_{,u})$ given by
 (\ref{xp}) and $\tilde{C}$ defined in (\ref{cagain}).
We correctly reproduce the following,
\begin{equation}
\tilde{C} = \pm e^{2A(u_c)}Q(u_c)
\end{equation}
and again choosing the positive sign we have,
\begin{equation}
\Pi_x^u = -\frac{e^{2A(u_c)}}{2\pi\alpha'}Q(u_c)v = f_{\mathsf{drag}}.
\end{equation}

We can now easily extend these arguments to a sheet-like defect. This comes down to setting $n=1$ and continuing to
demand that $m=n$. Since our general solution to (\ref{uccond}) does not depend on the dimensionality of the defect, we will again find
the same cutoff velocity for the sheet.  This supports the interpretation of \cite{superfluid} as this cutoff velocity is a property of
the system and not of the defect. The drag force will be modified as it is proportional to $\tilde{C}$,
which depends on $n$ through (\ref{cagain}).  We find,
\begin{equation}
\tilde{C}_{\mathsf{sheet}} = \pm e^{3A(u_c)}Q(u_c)
\end{equation}
and
\begin{equation}
\Pi_x^u = -\frac{e^{3A(u_c)}}{2\pi\alpha'}Q(u_c)v = f_{\mathsf{sheet},\mathsf{drag}}
\end{equation}

Like in the case of a dragging string in this holographic superfluid background, our analysis
has been performed entirely in the effective four dimensional language. While the background
itself is a consistent truncation of a full ten dimensional solution, it is not entirely obvious
what sort of object is described by the defect action eq. (\ref{defectaction}) with
the specific form fo $Q(\eta)$ from the
ten 10 dimensional point of view.
For the case of dragging strings in the background of five-dimensional charged black
holes that correspond to spinning black branes in ten dimension this question
has been carefully addressed in \cite{Herzog:2007kh} and indeed the use of the analog of
eq. (\ref{defectaction}) turned out to be questionable in that case. Here we take the
point of view of simply being interested in an effective four dimensional description and
take the action of the form eq. (\ref{defectaction}) 
as it is the most general two derivative
action consistent 
with symmetries.


\section{Discussion}

We gave a systematic study of dragging sheets in arbitrary holographic metrics. 
Our results reconfirm in this most general setting the general structure that was found for pointlike defects
 in general holographic metrics as well as for the study of dragging sheets in anti-de Sitter spaces: 
 the energy loss is completely insensitive to microscopic details of the system and only depends on the 
 velocity via an overall blueshifted energy density. This seems to be the most general characteristic of energy loss 
 at ``strong coupling", where a particle interpretation of the medium is not possible.

An example that may have a real world counterpart is the study of string like defects
 (corresponding to dragging membranes) in holographic superconductors. 
 Vortices in superfluid Helium and their energy loss can be studied experimentally. 
 To the extent that holographic superfluids and superconductors are candidates for real world systems,
  the loss rate experienced by a vortex in such a medium could be a physical observable.


\section*{Acknowledgments}
We'd like to thank Christopher Herzog and Stefan Janiszewski for useful discussions. This work was funded in part by the U.S. DOE under grant No. DE-FG02-96ER40956.

%
\bibliographystyle{apsrev} 
\bibliography{ref} 

\begin{thebibliography}{19}
\expandafter\ifx\csname natexlab\endcsname\relax\def\natexlab#1{#1}\fi
\expandafter\ifx\csname bibnamefont\endcsname\relax
  \def\bibnamefont#1{#1}\fi
\expandafter\ifx\csname bibfnamefont\endcsname\relax
  \def\bibfnamefont#1{#1}\fi
\expandafter\ifx\csname citenamefont\endcsname\relax
  \def\citenamefont#1{#1}\fi
\expandafter\ifx\csname url\endcsname\relax
  \def\url#1{\texttt{#1}}\fi
\expandafter\ifx\csname urlprefix\endcsname\relax\def\urlprefix{URL }\fi
\providecommand{\bibinfo}[2]{#2}
\providecommand{\eprint}[2][]{\url{#2}}

\bibitem[{\citenamefont{Maldacena}(1998)}]{Maldacena:1997re}
\bibinfo{author}{\bibfnamefont{J.~M.} \bibnamefont{Maldacena}},
  \bibinfo{journal}{Adv.Theor.Math.Phys.} \textbf{\bibinfo{volume}{2}},
  \bibinfo{pages}{231} (\bibinfo{year}{1998}), \eprint{hep-th/9711200}.

\bibitem[{\citenamefont{Gubser et~al.}(1998)\citenamefont{Gubser, Klebanov, and
  Polyakov}}]{Gubser:1998bc}
\bibinfo{author}{\bibfnamefont{S.}~\bibnamefont{Gubser}},
  \bibinfo{author}{\bibfnamefont{I.~R.} \bibnamefont{Klebanov}},
  \bibnamefont{and} \bibinfo{author}{\bibfnamefont{A.~M.}
  \bibnamefont{Polyakov}}, \bibinfo{journal}{Phys.Lett.}
  \textbf{\bibinfo{volume}{B428}}, \bibinfo{pages}{105} (\bibinfo{year}{1998}),
  \eprint{hep-th/9802109}.

\bibitem[{\citenamefont{Witten}(1998)}]{Witten:1998qj}
\bibinfo{author}{\bibfnamefont{E.}~\bibnamefont{Witten}},
  \bibinfo{journal}{Adv.Theor.Math.Phys.} \textbf{\bibinfo{volume}{2}},
  \bibinfo{pages}{253} (\bibinfo{year}{1998}), \eprint{hep-th/9802150}.

\bibitem[{\citenamefont{Karch}(2011)}]{Karch:2011ds}
\bibinfo{author}{\bibfnamefont{A.}~\bibnamefont{Karch}} (\bibinfo{year}{2011}),
  \eprint{1108.4014}.

\bibitem[{\citenamefont{Janiszewski and Karch}(2011)}]{stefan}
\bibinfo{author}{\bibfnamefont{S.}~\bibnamefont{Janiszewski}} \bibnamefont{and}
  \bibinfo{author}{\bibfnamefont{A.}~\bibnamefont{Karch}},
  \bibinfo{journal}{JHEP} \textbf{\bibinfo{volume}{1111}}, \bibinfo{pages}{044}
  (\bibinfo{year}{2011}), \eprint{1106.4010}.

\bibitem[{\citenamefont{Herzog et~al.}(2006)\citenamefont{Herzog, Karch,
  Kovtun, Kozcaz, and Yaffe}}]{karchquarkdrag}
\bibinfo{author}{\bibfnamefont{C.~P.} \bibnamefont{Herzog}},
  \bibinfo{author}{\bibfnamefont{A.}~\bibnamefont{Karch}},
  \bibinfo{author}{\bibfnamefont{P.}~\bibnamefont{Kovtun}},
  \bibinfo{author}{\bibfnamefont{C.}~\bibnamefont{Kozcaz}}, \bibnamefont{and}
  \bibinfo{author}{\bibfnamefont{L.~G.} \bibnamefont{Yaffe}},
  \bibinfo{journal}{JHEP} \textbf{\bibinfo{volume}{07}}, \bibinfo{pages}{013}
  (\bibinfo{year}{2006}), \eprint{hep-th/0605158}.

\bibitem[{\citenamefont{Herzog}(2006)}]{herzogquarkdrag}
\bibinfo{author}{\bibfnamefont{C.~P.} \bibnamefont{Herzog}},
  \bibinfo{journal}{JHEP} \textbf{\bibinfo{volume}{09}}, \bibinfo{pages}{032}
  (\bibinfo{year}{2006}), \eprint{hep-th/0605191}.

\bibitem[{\citenamefont{Janiszewski}(2011)}]{Janiszewski:2011sx}
\bibinfo{author}{\bibfnamefont{S.}~\bibnamefont{Janiszewski}}
  (\bibinfo{year}{2011}), \bibinfo{note}{* Temporary entry *},
  \eprint{1112.0085}.

\bibitem[{\citenamefont{Karch and Katz}(2002)}]{Karch:2002sh}
\bibinfo{author}{\bibfnamefont{A.}~\bibnamefont{Karch}} \bibnamefont{and}
  \bibinfo{author}{\bibfnamefont{E.}~\bibnamefont{Katz}},
  \bibinfo{journal}{JHEP} \textbf{\bibinfo{volume}{0206}}, \bibinfo{pages}{043}
  (\bibinfo{year}{2002}), \eprint{hep-th/0205236}.

\bibitem[{\citenamefont{Breitenlohner and
  Freedman}(1982)}]{Breitenlohner:1982bm}
\bibinfo{author}{\bibfnamefont{P.}~\bibnamefont{Breitenlohner}}
  \bibnamefont{and} \bibinfo{author}{\bibfnamefont{D.~Z.}
  \bibnamefont{Freedman}}, \bibinfo{journal}{Phys.Lett.}
  \textbf{\bibinfo{volume}{B115}}, \bibinfo{pages}{197} (\bibinfo{year}{1982}).

\bibitem[{\citenamefont{Camino et~al.}(2001)\citenamefont{Camino, Paredes, and
  Ramallo}}]{Camino:2001at}
\bibinfo{author}{\bibfnamefont{J.}~\bibnamefont{Camino}},
  \bibinfo{author}{\bibfnamefont{A.}~\bibnamefont{Paredes}}, \bibnamefont{and}
  \bibinfo{author}{\bibfnamefont{A.}~\bibnamefont{Ramallo}},
  \bibinfo{journal}{JHEP} \textbf{\bibinfo{volume}{0105}}, \bibinfo{pages}{011}
  (\bibinfo{year}{2001}), \eprint{hep-th/0104082}.

\bibitem[{\citenamefont{Itzhaki et~al.}(1998)\citenamefont{Itzhaki, Maldacena,
  Sonnenschein, and Yankielowicz}}]{pbranes}
\bibinfo{author}{\bibfnamefont{N.}~\bibnamefont{Itzhaki}},
  \bibinfo{author}{\bibfnamefont{J.~M.} \bibnamefont{Maldacena}},
  \bibinfo{author}{\bibfnamefont{J.}~\bibnamefont{Sonnenschein}},
  \bibnamefont{and}
  \bibinfo{author}{\bibfnamefont{S.}~\bibnamefont{Yankielowicz}},
  \bibinfo{journal}{Phys.Rev.} \textbf{\bibinfo{volume}{D58}},
  \bibinfo{pages}{046004} (\bibinfo{year}{1998}), \eprint{hep-th/9802042}.

\bibitem[{\citenamefont{Kanitscheider and
  Skenderis}(2009)}]{Kanitscheider:2009as}
\bibinfo{author}{\bibfnamefont{I.}~\bibnamefont{Kanitscheider}}
  \bibnamefont{and}
  \bibinfo{author}{\bibfnamefont{K.}~\bibnamefont{Skenderis}},
  \bibinfo{journal}{JHEP} \textbf{\bibinfo{volume}{0904}}, \bibinfo{pages}{062}
  (\bibinfo{year}{2009}), \eprint{0901.1487}.

\bibitem[{\citenamefont{Karch et~al.}(2009)\citenamefont{Karch, Kulaxizi, and
  Parnachev}}]{Karch:2009eb}
\bibinfo{author}{\bibfnamefont{A.}~\bibnamefont{Karch}},
  \bibinfo{author}{\bibfnamefont{M.}~\bibnamefont{Kulaxizi}}, \bibnamefont{and}
  \bibinfo{author}{\bibfnamefont{A.}~\bibnamefont{Parnachev}},
  \bibinfo{journal}{JHEP} \textbf{\bibinfo{volume}{0911}}, \bibinfo{pages}{017}
  (\bibinfo{year}{2009}), \eprint{0908.3493}.

\bibitem[{\citenamefont{Gubser}(2008)}]{Gubser:2008px}
\bibinfo{author}{\bibfnamefont{S.~S.} \bibnamefont{Gubser}},
  \bibinfo{journal}{Phys.Rev.} \textbf{\bibinfo{volume}{D78}},
  \bibinfo{pages}{065034} (\bibinfo{year}{2008}), \eprint{0801.2977}.

\bibitem[{\citenamefont{Hartnoll et~al.}(2008)\citenamefont{Hartnoll, Herzog,
  and Horowitz}}]{Hartnoll:2008vx}
\bibinfo{author}{\bibfnamefont{S.~A.} \bibnamefont{Hartnoll}},
  \bibinfo{author}{\bibfnamefont{C.~P.} \bibnamefont{Herzog}},
  \bibnamefont{and} \bibinfo{author}{\bibfnamefont{G.~T.}
  \bibnamefont{Horowitz}}, \bibinfo{journal}{Phys.Rev.Lett.}
  \textbf{\bibinfo{volume}{101}}, \bibinfo{pages}{031601}
  (\bibinfo{year}{2008}), \eprint{0803.3295}.

\bibitem[{\citenamefont{Gubser and Yarom}(2010)}]{superfluid}
\bibinfo{author}{\bibfnamefont{S.~S.} \bibnamefont{Gubser}} \bibnamefont{and}
  \bibinfo{author}{\bibfnamefont{A.}~\bibnamefont{Yarom}},
  \bibinfo{journal}{JHEP} \textbf{\bibinfo{volume}{1003}}, \bibinfo{pages}{041}
  (\bibinfo{year}{2010}), \eprint{0908.1392}.

\bibitem[{\citenamefont{Gubser et~al.}(2009)\citenamefont{Gubser, Herzog, Pufu,
  and Tesileanu}}]{Gubser:2009qm}
\bibinfo{author}{\bibfnamefont{S.~S.} \bibnamefont{Gubser}},
  \bibinfo{author}{\bibfnamefont{C.~P.} \bibnamefont{Herzog}},
  \bibinfo{author}{\bibfnamefont{S.~S.} \bibnamefont{Pufu}}, \bibnamefont{and}
  \bibinfo{author}{\bibfnamefont{T.}~\bibnamefont{Tesileanu}},
  \bibinfo{journal}{Phys. Rev. Lett.} \textbf{\bibinfo{volume}{103}},
  \bibinfo{pages}{141601} (\bibinfo{year}{2009}), \eprint{0907.3510}.

\bibitem[{\citenamefont{Herzog and Vuorinen}(2007)}]{Herzog:2007kh}
\bibinfo{author}{\bibfnamefont{C.}~\bibnamefont{Herzog}} \bibnamefont{and}
  \bibinfo{author}{\bibfnamefont{A.}~\bibnamefont{Vuorinen}},
  \bibinfo{journal}{JHEP} \textbf{\bibinfo{volume}{0710}}, \bibinfo{pages}{087}
  (\bibinfo{year}{2007}), \eprint{0708.0609}.

\end{thebibliography}

\end{document}